\documentclass[aps,reprint,superscriptaddress,amsmath,amssymb]{revtex4-2}

\usepackage{graphicx}
\usepackage{dcolumn}
\usepackage{bm}
\usepackage{hyperref}

\usepackage{xcolor}

\begin{document}

\preprint{APS/123-QED}

\title{
Unexpected increase of intensity-dependent excitonic second- and third-harmonic generation induced by static electric fields}


\author{Ruixin Zuo}
\affiliation{
 Department of Physics, Paderborn University, Warburger Strasse 100, D-33098 Paderborn, Germany 
}
\author{Matthias Reichelt}
\affiliation{
 Department of Physics, Paderborn University, Warburger Strasse 100, D-33098 Paderborn, Germany 
}
\author{Cong Ngo}
\affiliation{
 Department of Physics, Paderborn University, Warburger Strasse 100, D-33098 Paderborn, Germany 
}
\author{Xiaohong Song}
\affiliation{
 School of Physics and Optoelectronic Engineering, Hainan University, Haikou 570288, China}%

\author{Weifeng Yang}
\affiliation{
 School of Physics and Optoelectronic Engineering, Hainan University, Haikou 570288, China}%
\affiliation{
Center for Theoretical Physics, Hainan University, Haikou 570288, China}%

\author{Torsten Meier}%
 \email{torsten.meier@uni-paderborn.de}
\affiliation{
 Department of Physics, Paderborn University, Warburger Strasse 100, D-33098 Paderborn, Germany 
}

\date{\today}

\begin{abstract}
We compute and analyze the dependence of excitonic second- and third-harmonic generation (SHG/THG) as a function of the optical excitation intensity in the presence of static electric fields by solving the semiconductor Bloch equations.
Our simulations are performed for excitation of the strongly bound intralayer exciton of an inversion-symmetric homobilayer of $MoS_2$ with in-plane electric fields.
We demonstrate that for resonant excitation at the 1s K-exciton the SHG and the THG show complex dependencies on both the strength of the static field and the peak amplitude of the optical pulse.
For sufficiently intense optical excitation, the THG increases and the SHG increases superlinearly with the amplitude of the static field as long as exciton ionization is not yet dominating.
Microscopic simulations demonstrate that these dependencies arise from an interplay between several effects including static and transient Stark shifts, exciton ionization, Wannier-Stark localization, off-resonant Rabi oscillations, and a modified interference between optical nonlinearities induced by the intraband acceleration.
Our findings offer several new possibilities for controlling the strong-field dynamics of systems with strongly bound excitons.
\end{abstract}
\maketitle


\section{\label{sec:Introduction}Introduction}

As is well known, homogeneous static electric fields are able to drastically change the electronic properties of crystals \cite{bloch28,zener34,wannier60,Wannier1962,franz,keldysh,thar}.
Modifications of the electron wave functions and the electronic dynamics by electrical biasing can be detected using suitable frequency- and time-domain optical techniques that excite and/or probe interband transitions.
Indeed, over the last several decades a number of intriguing effects induced by static electric fields have been observed in semiconductors and semiconductor nanostructures like quantum wells and superlattices, see, e.g.,
\cite{FKexp1,FKexp2,FKexp3,Mendez1988,Voisin1988,FKexp4SL,FKexp5SL,BO1,BO2,BOwp,Xion}. 
For example, field-induced tunneling of the electron wave function into the band gap and the accompanying oscillatory behavior above the band gap, known as the Franz-Keldysh (FK) effect \cite{franz,keldysh,thar}, is visible in optical absorption spectra as
finite absorption below the band gap and above-band gap oscillations \cite{FKexp1,FKexp2,FKexp3,FKexp4SL,FKexp5SL}. Furthermore, electric fields localize the electronic wave functions, which are extended Bloch functions in the absence of a bias, in the direction in which they are applied. This Wannier-Stark (WS) localization \cite{wannier60,Wannier1962} is visible in optical spectra when the field strength is sufficiently strong such that 
the separation between WS ladder states exceeds the optical linewidth and transitions to individual WS states can be resolved \cite{Mendez1988,Voisin1988}. To be able to observe the time-domain counterpart of WS localization, i.e., Bloch oscillations (BO) \cite{bloch28,zener34}, the period for one BO must be shorter than the dephasing times. Since the dielectric breakdown limits the amplitudes of static electric fields that can maximally be applied without destroying the crystal, WS localization \cite{Mendez1988,Voisin1988} and BOs \cite{BO1,BO2} were first detected in semiconductor superlattices which are artificial crystals with a unit cell that is much larger than lattice constants of natural crystals. 
As the splitting between WS states is proportional to the length of the unit cell and the BO period is inversely proportional to the length of the unit cell, the requirements to observe these phenomena could be reached in semiconductor superlattices for field strengths well below the dielectric breakdown fields.

When semiconductors are excited off-resonantly far below the band gap with ultrashort light pulses, the peak electric field amplitudes (typically several $MV/cm$ up to some 10~$MV/cm$) can significantly exceed the breakdown fields without destroying the material. This regime is nowadays routinely reached in many strong-field experiments and is, e.g., utilized for the generation of high harmonics from semiconductors \cite{HHG_0,dynBO,Luu2015,Hohenleutner2015,Ndabashimiye2016,Luu2018,HHG_rev1,HHG_rev2,Heide2022,Heide2024,HHG_rev3,Zhang2025,Kira2025}. 
In such experiments the fields are often so strong that transient BOs, i.e., a time-dependent motion of the wave vector in field direction through the entire Brillouin zone during one cycle of the exciting optical field, take place during the excitation \cite{dynBO}. Thereby such highly-intense pulses strongly modify the electronic properties on ultrashort time scales and can, e.g., realize transient WS localization which leads to distinct modifications of optical spectra \cite{Schmidt2018}.

Due to the Coulomb attraction between photoexcited electrons and holes, excitonic resonances, representing bound electron-hole states with typically hydrogen-like wave functions, appear as discrete absorption lines below the band gap \cite{Haug,Kalt}. The binding energy of excitons in semiconductors and semiconductor nanostructures depends strongly on the material and, in particular, on the dimensionality and varies from a few $meV$ up to about $100~meV$.
The wave function of the lowest 1s exciton is typically, at least to very good approximation, symmetric in space and therefore a static electric field will typically result in a quadratic (static or dc) Stark shift to lower energies \cite{X_Stark1,X_ion1,X_Stark2}. 
Since an electric field corresponds to a potential that varies linearly in space and thus tilts the attractive Coulomb potential, it also leads to a finite probability that electrons and holes separate from each other, i.e., that the exciton ionizes. The reduced lifetime of the exciton due to ionization can be observed as a field-dependent broadening of the exciton peak in optical spectra \cite{X_ion1,X_ion2} or as a more rapid decay in time-domain spectroscopy \cite{Xion}.
The situation changes for sufficiently strong static fields that lead to WS localization. In this regime the states in the field direction are localized and the Coulomb attraction couples the electron and hole WS states to form WS excitons \cite{Dignam1990,Linder1996}.

For sufficiently intense resonant excitation, Rabi flopping of semiconductor excitons can be achieved \cite{rabi_x_1,rabi_x_2}. Excitonic Rabi flopping differs distinctively from what one would expect for a simple two-level system and is significantly influenced by many-body correlations.
When excitonic systems are excited off-resonantly below the resonance by intense fields  off-resonant Rabi oscillations, i.e, an incomplete oscillation of the occupation and the optical polarization, can be induced \cite{alleneberly,offres1}. They can be viewed as a kind of partial Rabi oscillations where no complete occupation of the upper state is realized, however, the polarization still changes sign during the dynamics.
Furthermore, when semiconductors and semiconductor nanostructures are excited weakly detuned below the resonance, excitons shift towards higher energies due to the optical Stark effect \cite{osa1,osa2,osa3}. The strength of this dynamical or ac Stark shift is proportional to the field intensity and inversely proportional to the detuning. 
In particular, when pulses with different polarizations are used, the excitonic Stark effect was shown to be strongly influenced by exciton-exciton interactions, i.e., many-body correlations \cite{Sieh1999}.
Furthermore, also the excitonic dynamical FK effect has been reported in semiconductor quantum wells where a frequency- and intensity-dependent shift of the excitonic resonance under intense THz irradiation has been observed \cite{Nordstrom1998}. 

The family of two-dimensional (2D) transition-metal dichalcogenides (TMD) semiconductors shows numerous fascinating physical properties resulting from their reduced dimensionality and crystal symmetry, which are distinct from their bulk counterparts
\cite{Wang2012, Xiao2012,  Xu2014, Manchon2015, Moody2015, Xu2016, SHG_bilayer_1, Selig2016, Liu2017, Wang2018,  Mak2018, Ribeiro2018, Langer2018, Mueller2018, Yoshikawa2019, Trovatello2020, SHG_bilayer_2, Helmrich2021, Ciarrocchi2022,   Kobayashi2023, Hader2023, Lee2024}. 
Due to the quantum confinement and decreased dielectric screening of the Coulomb interaction, excitons in 2D TMD are observable even at room temperature and dominate the optical properties. 2D TMD host exceptionally high exciton binding energies of several $100~meV$ and hence are ideally suited for studying exciton physics. 

Due to the large exciton binding energies and the associated small excitonic Bohr radii, the amplitudes of static fields that induce significant changes in the excitonic properties are one to two orders of magnitude higher than those for ordinary semiconductors and semiconductor nanostructures. Sufficiently strong field amplitudes lead to measurable excitonic Stark shifts which arise from the coupling to other exciton states and exciton ionization \cite{Scharf2016,Pedersen2016,Massicotte2018,Zhu2023}. 

The in-plane breakdown fields of TMDs are typically on the order of $MV/cm$. As we show below, for such field amplitudes the electronic properties are significantly modified via WS localization. The very high absorption of TMD excitons leads also to sizeable optical Stark shifts and off-resonant Rabi oscillations for excitation with far off-resonant optical fields with peak amplitudes in the range of a few $MV/cm$. Thus the strong-field physics of excitons in the presence of optical and static fields is influenced by an interesting interplay of several phenomena.
As several times higher field amplitudes than the breakdown field can be used for pulsed excitation it is possible to investigate in TMDs ultrafast dynamical excitonic processes in highly non-perturbative regimes which are not reachable in most other systems. 

In this paper, we present and discuss the results of microscopic simulations for resonant excitonic second- and third-harmonic generation (SHG/THG). As a model system we investigate the strongly bound intralayer excitons of an inversion-symmetric homobilayer of $MoS_2$. We consider in-plane polarized optical pulses which are two- or three-photon resonant with the 1s K-exciton and study the dependencies of SHG and THG on the excitation intensity and on the amplitude of a homogeneous static in-plane electric field. Our results are obtained by numerical solutions of the semiconductor Bloch equations (SBE) including the many-body Coulomb interaction in time-dependent Hartree-Fock approximation and inter- and intraband excitations in the length gauge \cite{Haug,Golde2008,NgO2023}. The band structure and matrix elements are computed by density-functional theory \cite{elk}.

Since the homobilayer of $MoS_2$ is inversion symmetric, SHG is only possible when the symmetry is broken which can be achieved in several different ways. For example one can use static electric fields \cite{EFISH_1,EFISH_2} which is known as electric-field-induced second-harmonic generation (EFISH), an effect that has been widely used in different settings and on various material systems, see, e.g., \cite{EFISH_rev}.
For static fields applied perpendicular to the layers remarkable switching and tunability of SHG 
from 2H stacked $MoS_2$ homobilayer  
was demonstrated \cite{SHG_bilayer_1} and the role of interlayer excitons mediating SHG has been investigated  \cite{SHG_bilayer_2}.

Here, we consider optical and static fields which are co-linearly polarized in the plane of the layers and show that the dependence of the excitonic SHG on the amplitude of the static field changes qualitatively with the intensity of the optical field. In particular, we find a regime in which the SHG increases superlinearly with the amplitude of the static field.
Our microscopic simulations clarify that a combination of several effects including static and transient Stark shifts, off-resonant Rabi oscillations, exciton ionization, and WS localization contribute to the obtained complex field dependencies.
Also the computed excitonic THG 
shows a very complex dependence on the amplitudes of the static and the optical fields.
For moderate optical excitation intensity it exhibits an unusual increase as function of the amplitude of the static field when exciton ionization is not yet dominant. Besides the effects mentioned above, for the static-field-enhanced THG additionally higher-order optical excitations representing exciton-to-continuum-to-exciton transition are relevant. Our analysis shows that the intraband acceleration modifies the interference between optical nonlinearities, leading to a more constructive interference in the presence of a static field.

The paper is structured as follows. Sec.~\ref{sec:theory} presents the modeling of the system and the theoretical description of the dynamics including the light-matter and the many-body Coulomb interactions. Some additional details are provided in the Appendix. In Sec.~\ref{sec:results} we present and discuss excitonic optical absorption spectra in bilayer $MoS_2$ with in-plane electric fields and the dependence of the excitonic SHG and THG on the strengths of the static and optical fields. 
To explain these findings we analyze how the intraband-acceleration induced by static and oscillating fields changes the excitonic absorption. Furthermore we investigate the influence of static and transient Stark shifts and of off-resonant Rabi oscillations on the dynamics.
In Sec.~\ref{sec:conclusions} we briefly summarize our main findings.

\section{Theoretical model}
\label{sec:theory}

Our model system is a 2H stacked homobilayer of $MoS_2$ which consists of two van der Waals bonded monolayers which are rotated by $180^{\circ}$ with respect to each other. In contrast to the non-equivalence of
the two valleys K and K$^\prime$ including a finite splitting of bands which results from the lifted spin
degeneracy present in a monolayer, the bands of the 2H homobilayer are spin-degenerate throughout the
Brillouin zone (BZ) and have equivalent K valleys. It was shown that homobilayer TMDs can be used to tune the valley circular dichroism in a controllable way, e.g., by applying an interlayer bias \cite{Xu2014, Lee2016, Korm2018, Gong2013, Wu2013}. Here, we concentrate on intralayer excitonic effects in homobilayer $MoS_2$ excited and biased by in-plane polarized electric fields and compute the dynamical optical response by numerically solving the SBE. 

We start from a many-body Hamiltonian that contains three terms \cite{Haug,Golde2008,NgO2023}
\begin{equation}
H=H_0+H_{LM}+H_{C} .
\end{equation}
The electronic band structure is described by
\begin{equation}
H_0=\sum_{\lambda\bm{k}}\epsilon_{\lambda\bm{k}}a^{\dagger}_{\lambda\bm{k}}a_{\lambda\bm{k}},
\label{eq2}
\end{equation}
where $\lambda$ combines the spin and band indices and $a^{\dagger}_{\lambda\bm{k}}$ ($a_{\lambda\bm{k}}$) is the creation (annihilation) operator of an electron with crystal momentum $\hbar\bm{k}$ in band $\lambda$ and $\epsilon_{\lambda\bm{k}}$ is the band structure. The eigenfunctions of the crystal Hamiltonian are Bloch functions $\Psi_{\lambda,\bm{k}}(\bm{r})=e^{i\bm{k}\cdot\bm{r}}u_{\lambda,\bm{k}}(\bm{r})$. Using the length gauge, the light-matter interaction is described in the dipole approximation by \cite{NgO2023,Vanderbilt2018,Aversa1995,Virk2007}
\begin{equation}
H_{LM}=- e \bm{E}(t)\cdot\sum_{\lambda\lambda'\bm{k}\bm{k}'}\bm{r}^{\lambda\lambda'}_{\bm{k}\bm{k}'}a^{\dagger}_{\lambda\bm{k}}a_{\lambda'\bm{k}'}.
\end{equation}
Here, $\bm{E}(t)$ is the electric field and
\begin{equation}
e \bm{r}^{\lambda\lambda'}_{\bm{k}\bm{k}'}=[\bm{d}^{\lambda\lambda'}_{\bm{k}} +i e \delta_{\lambda\lambda'}\nabla_{\bm{k}}]\delta({\bm{k}-\bm{k}'})
\end{equation}
are the matrix elements of the position  operator
multiplied by the electron charge $e$, where $\bm{d}^{\lambda\lambda'}_{\bm{k}}=e\int d^3\bm{r}u_{\lambda,\bm{k}}(\bm{r})\nabla_{\bm{k}}u_{\lambda',\bm{k}}(\bm{r})$ denotes for $\lambda \neq \lambda'$ the interband dipole matrix elements and for $\lambda = \lambda'$ the Berry connection $\bm{d}^{\lambda\lambda}_{\bm{k}}/e$ and $\nabla_{\bm{k}}$ describes the intraband acceleration by the electric field. The many-body Coulomb interaction between photoexcited carriers reads
\begin{equation}
H_{C}=\frac{1}{2}\sum\limits_{\substack{\lambda\lambda' \bm{k}\bm{k}' \bm{q} \\\bm{q} \neq  0}}V_{\bm{q},\bm{k},\bm{k}'}^{\lambda\lambda'}a^{\dagger}_{\lambda\bm{k}}a^{\dagger}_{\lambda'\bm{k}'}a_{\lambda'\bm{k}'+\bm{q}}a_{\lambda\bm{k}-\bm{q}}
\end{equation}
with the quasi-2D Coulomb matrix element
\begin{eqnarray}   
V_{\bm{q},\bm{k},\bm{k}'}^{\lambda\lambda'}=&& \int d^3\bm{r}\int d^3\bm{r}'u^{*}_{\lambda\bm{k}}(\bm{r})u^{*}_{\lambda'\bm{k}'}(\bm{r}')\nonumber\\
&&\times W_{\bm{q}}(z,z')u_{\lambda'\bm{k}'+\bm{q}}(\bm{r}')u_{\lambda \bm{k}-\bm{q}}(\bm{r}),
\end{eqnarray}
where $W_{\bm{q}}(z,z')$ is the Coulomb potential including dielectric screening that is described by a dielectric function which is derived from Poisson's equation \cite{Meckbach2018}; see Appendix~\ref{sec:ApdB} for more details. The Coulomb interaction leads to renormalizations of the single-particle band structure and the Rabi frequency. The latter results in the formation of bound electron-hole pairs, i.e., excitons, which strongly modify the optical properties.

The dynamics of the photoexcited carriers is analyzed using the Heisenberg equations of motion for the microscopic interband polarizations $p^{\lambda\lambda'}_{\bm{k}}=\langle a^{\dagger}_{\lambda'\bm{k}}a_{\lambda\bm{k}}\rangle$ and the carrier occupations $n^{\lambda}_{\bm{k}}=\langle a^{\dagger}_{\lambda\bm{k}}a_{\lambda\bm{k}}\rangle$. The resulting coupled equations of motion are known as the semiconductor Bloch equations (SBE) which in time-dependent Hartree-Fock approximation read  \cite{Haug,Golde2008,NgO2023}
\begin{subequations}
\begin{eqnarray}
\frac{\partial}{\partial t}p^{\lambda\lambda'}_{\bm{k}}=&&-\frac{i}{\hbar}(\varepsilon^{\lambda}_{\bm{k}}-\varepsilon^{\lambda'}_{\bm{k}})p^{\lambda\lambda'}_{\bm{k}}+\frac{e}{\hbar}\bm{E}(t)\cdot\nabla_{\bm{k}}p^{\lambda\lambda'}_{\bm{k}}
\nonumber\\
&&+\frac{i}{\hbar}\Omega^{\lambda\lambda'}_{\bm{k}}(n^{\lambda'}_{\bm{k}}- n^{\lambda}_{\bm{k}})
- \frac{p^{\lambda\lambda'}_{\bm{k}}}{T_2} ,
\label{subeq:7a}
\\
\frac{\partial}{\partial t}n^{\lambda}_{\bm{k}}=&&\frac{2}{\hbar}Im[\sum_{\lambda'\neq\lambda}{\Omega^{\lambda\lambda'}_{\bm{k}}}^{*}p^{\lambda\lambda'}_{\bm{k}}]\nonumber+\frac{e}{\hbar}\bm{E}(t)\cdot\nabla_{\bm{k}}n^{\lambda}_{\bm{k}},
\label{subeq:7b}
\\
\end{eqnarray}
\end{subequations}
where 
\begin{equation}
    \Omega^{\lambda\lambda'}_{\bm{k}}=\bm{d}^{\lambda\lambda'}_{\bm{k}}\cdot \bm{E}(t)+\sum_{\bm{k}'}V^{\lambda\lambda'}_{\bm{k}-\bm{k}',\bm{k},\bm{k}'}p^{\lambda\lambda'}_{\bm{k}'}
\end{equation}
is the renormalized Rabi frequency,
\begin{equation}
\varepsilon^{\lambda}_{\bm{k}}=\epsilon^{\lambda}_{\bm{k}}-\sum_{\bm{k}'}V^{\lambda\lambda}_{\bm{k}-\bm{k}',\bm{k},\bm{k}'}n^{\lambda}_{\bm{k}'}
\end{equation}
is the renormalized energy, and $T_2$ is the dephasing time of the microscopic polarizations.

The light field emitted by a coherently excited semiconductor contains two sources and is described by $E_{out}(t) \propto \frac{\partial^2}{\partial t^2}\bm{P}(t)+\frac{\partial}{\partial t}\bm{J}(t)$ \cite{Golde2008,Huttner2017} with the macroscopic polarization $\bm{P}(t)$ originating from interband transitions
\begin{equation}
\bm{P}(t)=\sum\limits_{\substack{\lambda\lambda' \\ \lambda \neq \lambda'}}
\int_{BZ}d\bm{k}\bm{d}^{\lambda\lambda'}_{\bm{k}}p^{\lambda\lambda'}_{\bm{k}}
\end{equation}
and the macroscopic charge current $\bm{J}(t)$ due to intraband currents
\begin{equation}
    \bm{J}(t)= e \sum_{\lambda}\int_{BZ}d\bm{k}\bm{v}^{\lambda}_{\bm{k}}n^{\lambda}_{\bm{k}}
\end{equation}
where the integration runs over the first BZ and $\bm{v}^{\lambda}_{\bm{k}}=\nabla_{\bm{k}}\varepsilon^{\lambda}_{\bm{k}}$ is the group velocity in band $\lambda$. 

We perform density functional calculations to obtain the ground state of the 2H homobilayer of $MoS_2$ using the Elk code \cite{elk} with the generalized gradient approximations (GGA). The band gap is adjusted using the scissors operator. The multiband twisted parallel transport gauge \cite{NgO2023,Vanderbilt2018,Thong2021,Gaarde_gauge} is implemented on the eigenfunctions resulting in matrix elements that vary smoothly as function of wave vector. For the homobilayer $MoS_2$ system, the spin-degenerate bands are formed by orbitals from different layers. Due to interlayer hopping which hybridizes the valence band states of different layers, the splitting of bands occurs with the increase of the number of layers. Particularly, the valence band splitting at the $\Gamma$ point is much larger than at the $K$ point for bilayer $MoS_2$ and therefore the band gap is indirect \cite{Liu2015}. However, the lowest direct interband transition energy still occurs at the K point with a band gap of $E_g=2.336~eV$. According to the optical selection rules \cite{Liu2015}, in each $K$ valley, the top valence band couples to the lowest conduction band via an intralayer transition and couples to another energetically close conduction band via an interlayer transition, the latter being rather weak. In the present work, we consider optical intralayer transitions via a four-band model (two degenerate conduction and two degenerate valence bands around the band gap). The peak strength of the interband dipole matrix elements for the two allowed transitions of opposite spins is $0.153~e~nm$. In Appendix~\ref{sec:ApdA}, we rewrite the SBE for one spin, i.e., a two-band model (the transition for the other spin has the same magnitude under in-plane polarized exciting fields) and also present a perturbative expansion of the SBE which is used for analyzing the dynamics for not too strong optical excitation.

In our numerical evaluations, we consider optical and static electric fields which are co-linearly polarized in the plane of the homobilayer, i.e., the total electric field is given by
$\bm{E}(t) = E(t)\bm{e}= ( E_{opt}(t) + E_{dc})\bm{e} $, where the unit vector $\bm{e}$ defines the polarization direction. Note that below we denote the peak amplitude of the optical field by $E_{opt}$.
Throughout this paper the fields are polarized in the zigzag direction.
We discuss the components of the interband polarization and the intraband current in this direction which allows us to ignore their vector characters. The optical pulses used in the numerical calculations have a Gaussian envelope with a full width at half maximum of 100~$fs$ and 
dephasing is modeled phenomenologically via a dephasing time of $T_2=50$~$fs$.\\

\section{\label{sec:results}Results and Discussion}

\begin{figure}[t]
\includegraphics[width=1.0\columnwidth]{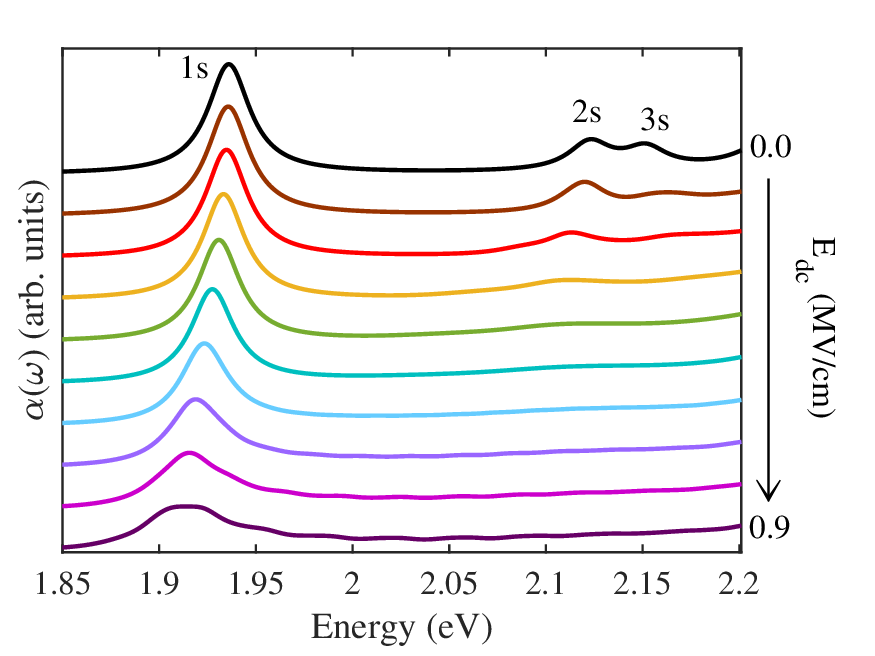}
\caption{Calculated linear absorption spectra for a freestanding 2H homobilayer $MoS_2$. With increasing static field amplitude $E_{dc}$ all excitons Stark-shift to lower energies. Simultaneously, the peak absorption is reduced and the absorption lines broaden due to exciton ionization. For $E_{dc}\ge0.8~MV/cm$ weak periodic oscillations appear which originate from WS localization. The curves are shifted vertically for clarity.}
\label{fig1}
\end{figure}

To determine the linear absorption, we solve Eq.~(\ref{subeq:7a}) for constant occupations using a weak ultrashort in-plane-polarized optical pulse. The spectral absorption is proportional to the imaginary part of the ratio between the Fourier-transformed macroscopic polarization and the optical electric field \cite{Haug}, i.e., 
\begin{equation}
    \alpha(\omega) \propto \omega Im [ P(\omega)/E_{opt}(\omega)]
\end{equation}
where the pulse spectrum $E_{opt}(\omega)$ is broad and covers the excitation of exciton and continuum states. Fig.~\ref{fig1} displays the linear absorption spectra of a homobilayer $MoS_2$ for different static electric fields $E_{dc}$ which induce a time-dependent change of the wave vectors according to the acceleration theorem. Without an electrical bias, the linear absorption spectrum provides the energetic positions of the three lowest s excitons which resemble a modified hydrogenic series. The position $E_{1s}=1.936~eV$ and the binding energy of about $400~meV$ of the 1s exciton are in reasonable agreement with reported experimental and theoretical data for layered TMDs \cite{Erben2018, Datta2022, Peimyoo2021}. 

\begin{figure}[t]
\includegraphics[width=1.0\columnwidth]{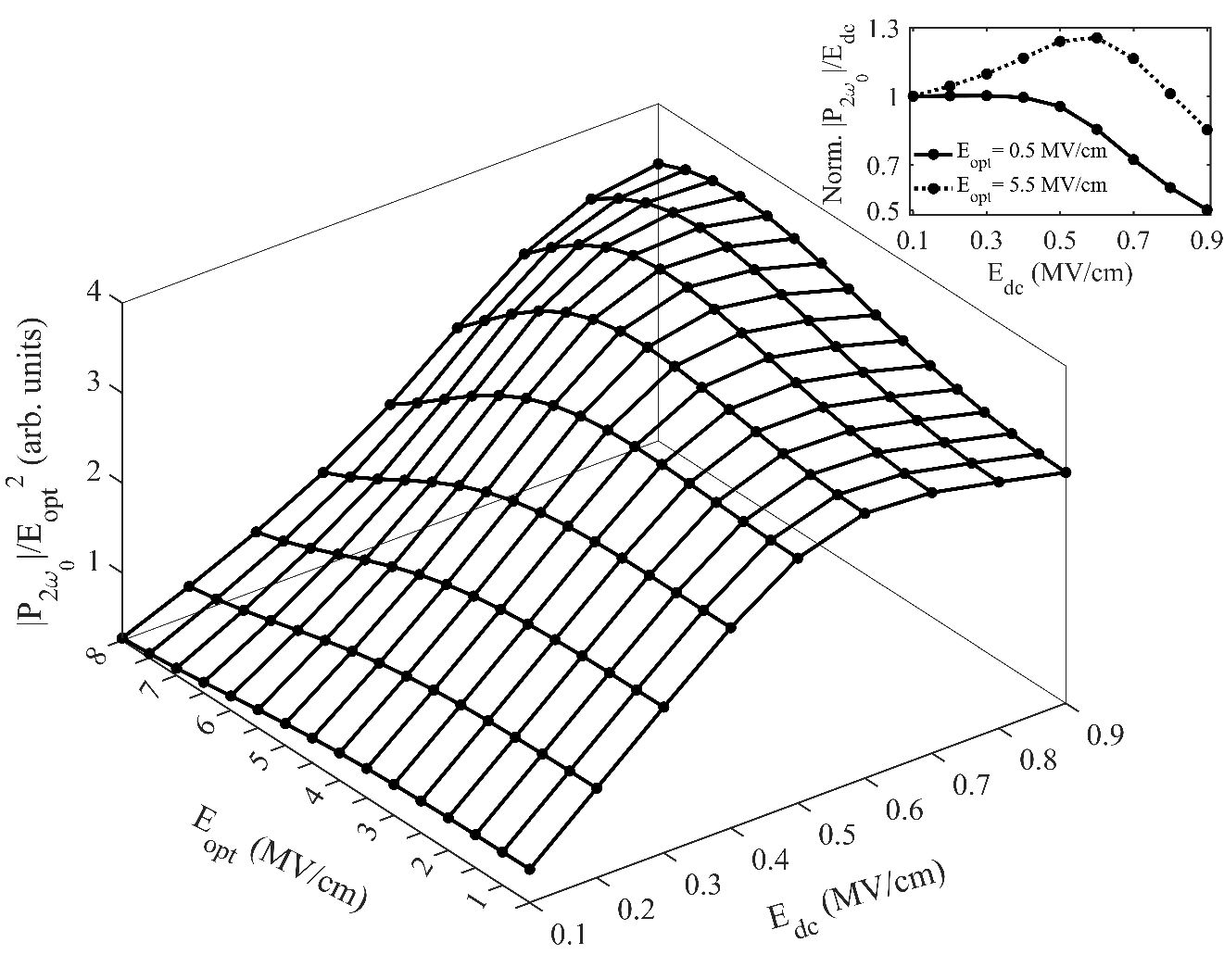}
\caption{Ratio between $|P_{2\omega_0}|$ and ${E_{opt}}^2$ with $|P_{2\omega_0}|=\int_{1.7\omega_0}^{2.3\omega_0}|P(\omega)|d\omega$ for various optical and static field amplitudes. The excitation frequency is $\hbar\omega_0=1.936~eV/2$. Inset: Normalized $|P_{2\omega_0}|/E_{dc}$ as a function of the static field amplitude for $E_{opt}=0.5~MV/cm$ (solid dotted line) and $E_{opt}=5.5~MV/cm$ (dashed dotted line). The symbols represent the calculated values and the lines are guides for the eye.}
\label{fig2}
\end{figure}

A static in-plane electric field breaks the spatial symmetry and induces a coupling among excitonic states of opposite symmetry, e.g., s and p excitons. The newly formed exciton states can be described as superpositions of the field-free exciton states. The energy of the lowest exciton shifts red with increasing field showing a quadratic dc Stark shift \cite{Zhu2023} which we denote by $\delta_{dc}$. Apart from the hybridization, a static field tilts the attractive Coulomb potential in real space which results in a finite probability that the electron and hole separate, i.e., that the exciton ionizes, which reduces the exciton lifetime with increasing field amplitude \cite{Kamban2021, Massicotte2018}. The dc Stark shift $\delta_{dc}$ and the broadening due to field-induced exciton ionization are clearly visible in Fig.~\ref{fig1}.
Due to their weaker binding energy which corresponds to a larger extension in real space, the higher excitons 3s and 2s are ionized already for significantly weaker fields than the 1s exciton. 
For $E_{dc}\geq 0.8~MV/cm$ weak periodic oscillations appear near and above the 1s exciton.
The energetic separation between adjacent peaks changes linearly with the static field amplitude and coincides with the expected splitting between different WS states, i.e., $e E_{dc} a$ where $a$ is the lattice constant. 
For $E_{dc}=0.8~MV/cm$, the splitting $e E_{dc} a$ is about $24~meV$ and is thus close  to the homogeneous linewidth which is already increased somewhat above its field-free values of $2 \hbar/T_2=26~meV$ due to exciton ionization.
This, as well as additional model simulations with a longer $T_2$ (not shown), confirms that the origin of these peaks is indeed WS localization.
For the considered strengths of the static field, the WS splitting is still much smaller than the exciton binding energy and therefore the WS localization distorts the exciton absorption rather weakly.

\begin{figure}[t]
\includegraphics[width=1.0\columnwidth]{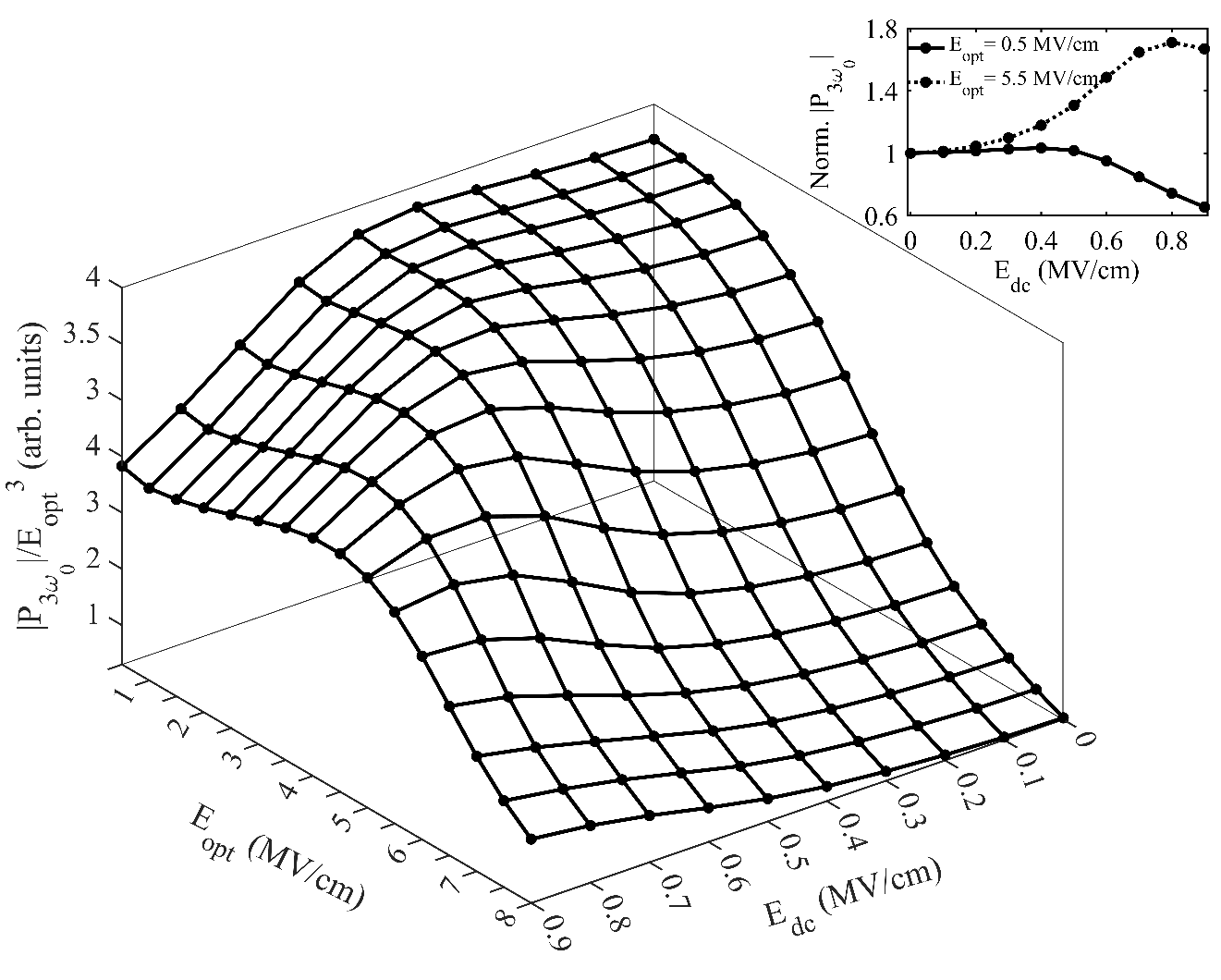}
\caption{Ratio between $|P_{3\omega_0}|$ and ${E_{opt}}^3$ with $|P_{3\omega_0}|=\int_{2.7\omega_0}^{3.3\omega_0}|P(\omega)|d\omega$ for various optical and static field amplitudes. The excitation frequency is $\hbar\omega_0=1.936~eV/3$. Inset: Normalized $|P_{3\omega_0}|$ as a function of static field amplitude for $E_{opt}=0.5~MV/cm$ (solid dotted line) and $E_{opt}=5.5~MV/cm$ (dashed dotted line). The symbols represent the calculated values and the lines are guides for the eye.}
\label{fig3}
\end{figure}

\begin{figure*}[t]
\includegraphics[width=2.0\columnwidth]{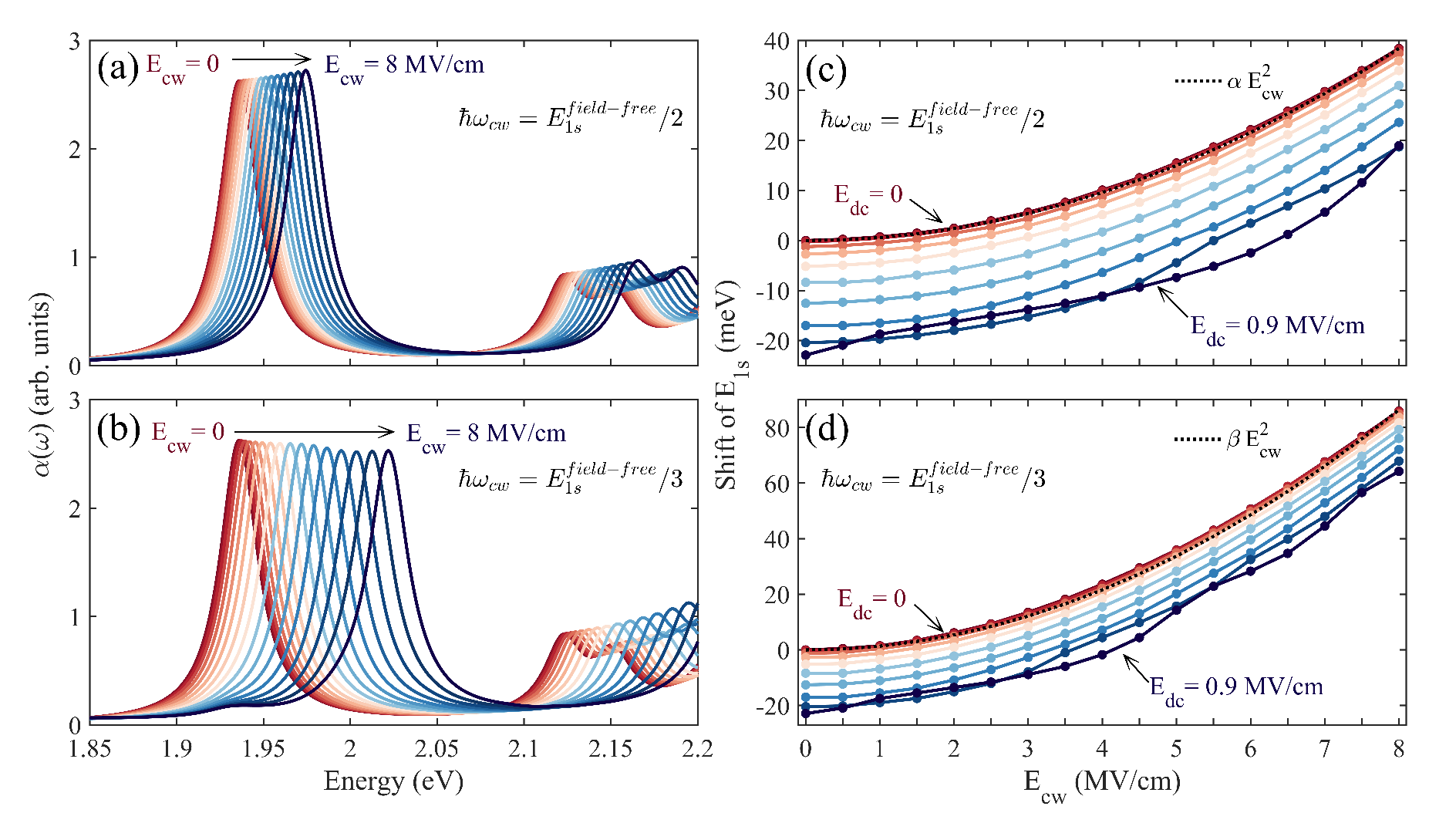}
\caption{Linear absorption spectra in the presence of a cw optical field with (a) $\hbar\omega_{cw}=1.936~eV/2$ and (b) $\hbar\omega_{cw}=1.936~eV/3$. (c) and (d) show the energetic shift of the main exciton resonance compared to the field-free position of the 1s exciton due to both static and cw optical fields for the cw fields used in (a) and (b), respectively. The oscillatory results for $E_{dc}\ge 0.7~MV/cm$ arise from WS localization. In this regime the largest excitonic WS peaks shows an oscillatory dependence on $E_{dc}$.}
\label{fig4}
\end{figure*}

We compute SHG and THG by solving the SBE, Eqs.~(\ref{subeq:7a}) and (\ref{subeq:7b}), for excitation pulses which are two- or three-photon resonant with the field-free 1s exciton, i.e., the photon energies are $\hbar \omega_0 = 1.936~eV/2$ or $\hbar \omega_0 = 1.936~eV/3$, respectively, for various static and optical field amplitudes.
The resulting SHG and THG shown in Figs.~\ref{fig2} and ~\ref{fig3} have been obtained by integrating $|P(\omega)|$ around $2 \omega_0$/$3 \omega_0$ and dividing the result by $E_{opt}^2$/$E_{opt}^3$, i.e., the dependence on the peak amplitude of the optical field in the perturbative limit.
For far off-resonant excitation the interplay between inter- and intraband transitions significantly influences the nonlinear optical response \cite{Schlaepfer2018, Nakagawa2022}. For the excitation conditions we consider here, the excited occupations are rather low and the resonantly excited excitonic polarization dominates. Therefore, the generated radiation is predominantly originating from the interband polarization whereas the contributions from the intraband currents are negligible. We verified that the resulting signal is reduced by a few orders of magnitude when the term $\bm{E}_{opt}(t)\cdot\nabla_{\bm{k}}p(\bm{k})$ in the SBE is artificially switched off. Therefore, beyond the linear order the SHG and THG are predominantly generated by intraband transitions. The dominant pathways to SHG and THG are that the linear polarizations $(p^{\lambda\lambda'}_{\bm{k}})^{(1)}$ are the sources of 
intraband transitions that generate the second-order and third-order polarizations, i.e.,
$(p^{\lambda\lambda'}_{\bm{k}})^{(2)}$ and $(p^{\lambda\lambda'}_{\bm{k}})^{(3)}$, which dominate the resonantly excited excitonic SHG and THG, respectively.

For a centrosymmetric system excited by a single-color pulse the generation of even harmonics is forbidden. Applying a static electric field breaks the symmetry and enables SHG which in the perturbative limit scales linearly with the field amplitude $E_{dc}$. As expected, for weak optical driving, i.e., small $E_{opt}$, the SHG shown in Fig.~\ref{fig2} starts to increase linearly with $E_{dc}$. Starting at about $E_{dc} \approx 0.5~MV/cm$ the increase gets weaker and turns into a decrease for $E_{dc} \ge 0.7~MV/cm$.
This dependence of the SHG on $E_{dc}$ for weak $E_{opt}$ originates from a combination of two effects which are visible in Fig.~\ref{fig1}. With increasing 
$E_{dc}$ the 1s exciton experiences a quadratic Stark shift $\delta_{dc}$ to lower energies such that the SHG is not excited fully resonantly. However, even if one would adjust for this Stark shift by tuning the excitation frequency accordingly, the SHG would still decrease for larger $E_{dc}$ since exciton ionization leads to a significant increase of the homogeneous linewidth, i.e., rapid dephasing.

For stronger optical excitation, i.e., larger $E_{opt}$, the SHG shows an unexpected dependence on $E_{dc}$. Above around $E_{opt}=4-5~MV/cm$ the optical driving leaves the perturbative limit, as can be seen by the reduction of the line for 
$E_{dc}=0.1~MV/cm$ for larger $E_{opt}$ in Fig.~\ref{fig2}. At the same time, the SHG starts to show a clear superlinear, approximately quadratic increase as function of $E_{dc}$, see also the inset of Fig.~\ref{fig2}, up to about $E_{dc}=0.5~MV/cm$. Furthermore, with increasing $E_{opt}$ the maximal SHG is reached for larger $E_{dc}$.

We now turn to the dependence of the THG on $E_{dc}$ and $E_{opt}$ shown in Fig.~\ref{fig3}. For weak optical driving the THG is unaffected by small bias fields, but it starts to decrease as function of $E_{dc}$ when the Stark shift $\delta_{dc}$ and exciton ionization become relevant.
For larger $E_{opt}$, when the optical excitation is beyond the perturbative limit, the THG shows an unexpected significant increase as function of $E_{dc}$ (note that we plot the amplitude of the polarization, i.e., the intensity is enhanced up to about a factor of three). As clearly shown in the inset of Fig.~\ref{fig3} for $E_{opt}=5.5~MV/cm$ the maximal THG is generated for dc field amplitudes at which for weaker optical driving exciton ionization already significantly reduces the THG. 
As will be analyzed below, static ($\delta_{dc}$) and dynamic optical Stark shifts ($\delta_{opt}(t)$), off-resonant Rabi oscillations, dc-field-induced exciton ionization and a partial stabilization in strong optical fields, WS localization, as well as a modified interference between higher-order optical nonlinearities are relevant in the investigated regime and contribute to the dependencies of the SHG and THG shown in Figs.~\ref{fig2}
and ~\ref{fig3}.

\begin{figure*}[t]
\includegraphics[width=2.0\columnwidth]{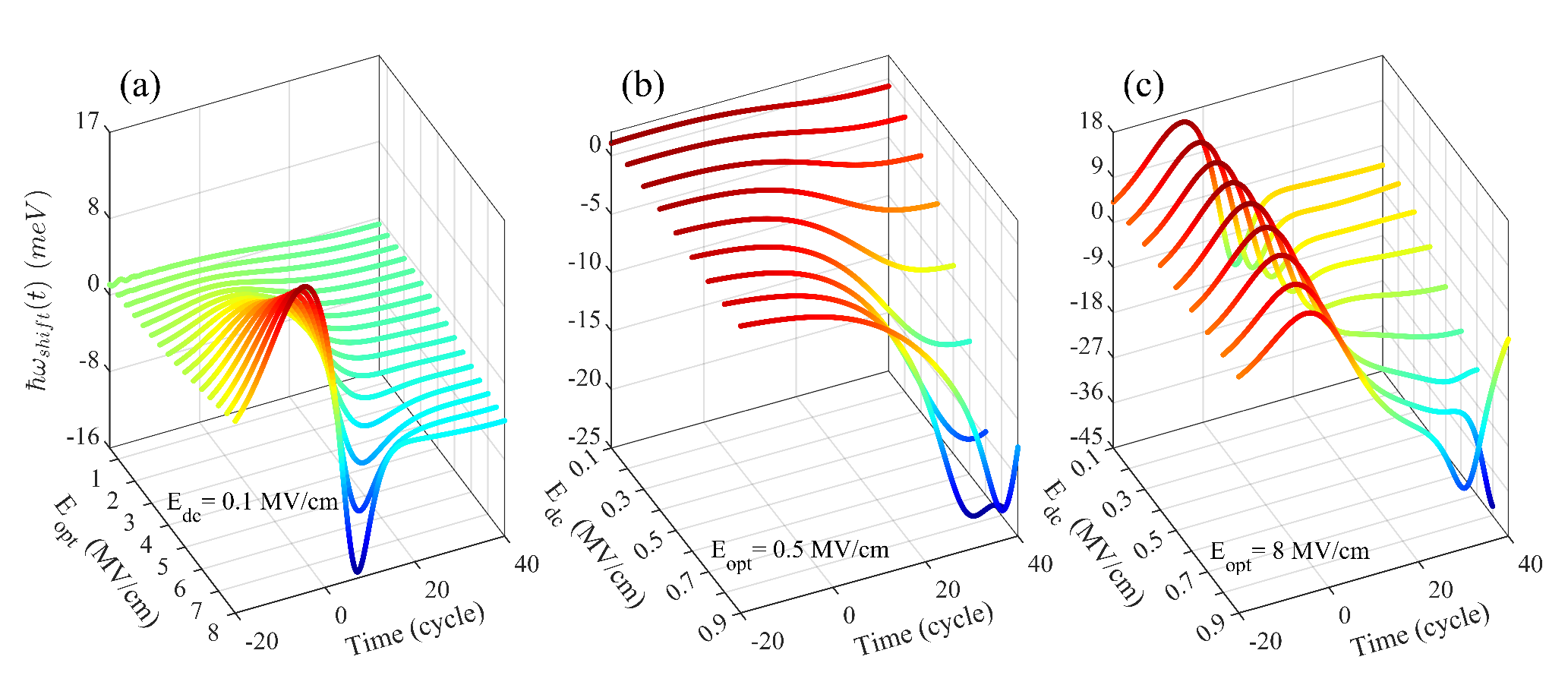}
\caption{$\hbar \omega_{shift}(t)$ of the SHG frequency for (a) varying the optical field amplitude with $E_{dc}=0.1~MV/cm$, (b) varying the static field amplitude with $E_{opt}=0.5~MV/cm$, and (c) the same as in (b) but for $E_{opt}=8.0~MV/cm$.
The excitation parameters are the same as those used in Fig.~\ref{fig2}.
The duration of an optical cycle is $4.27~fs$.}
\label{fig5}
\end{figure*}

To gain further insight into the change of the excitonic properties under the action of an oscillating field, we calculate the linear optical absorption in the presence of a constantly oscillating field, i.e., a continuous wave (cw) field, which is considered only in the intraband acceleration of Eq.~(\ref{subeq:7a}).
The linear absorption spectra for frequencies corresponding to the SHG and THG calculations and different field amplitudes are shown in Fig.~\ref{fig4}(a) and (b).
We clearly see that the peak of the 1s exciton shifts towards higher energies with increasing amplitude of the cw field, i.e., an optical Stark shift.
This shift can be understood to arise from the intraband acceleration of electrons and holes induced by the optical cw field. The averaging over the periodic dynamic field-induced change of the interband transition energy shifts the effective gap of the interband absorption to higher energies. The effective gap increases by the ponderomotive energy $U_p\propto {E_{cw}^2} / {\omega_{cw}^2}$ \cite{Kruchinin2018}.

The 1s exciton peaks seen in Fig.~\ref{fig4}(a) and (b) basically follow the shift of the band gap to higher energies which is quadratic in $E_{cw}$ and due to the smaller $\omega_{cw}$ larger for the THG than for the SHG excitation conditions.
Unlike the suppression of the absorption peak and the spectral broadening in the presence of a static field, the shape of the absorption peak in Fig.~\ref{fig4}(a) is nearly unchanged by the cw field. For lower frequencies of the cw field, exciton ionization can contribute to a certain extent which explains why the absorption peak is slightly reduced with increasing cw field amplitude for $\hbar\omega_0=1.936~eV/3$ shown in Fig.~\ref{fig4}(b).
If we carefully determine the exciton binding energy by taking the difference of the exciton peak to the effective band gap in the presence of the cw field, we find a small increase as function of $E_{cw}$. 
This can be understood to arise from the reduced band width in the polarization direction of the cw field 
which originates from the averaging of the transition energy over the oscillations of the wave vector induced by the optical field.
As a result the effective mass near the band gap increases, i.e., a stronger anisotropy of the band structure and thus of the exciton evolves. This effect can also be viewed as the onset of dynamical localization \cite{dunlapkenkre86prb,holthaus92prl} which was shown to lead to an increased anisotropy or even a change of the effective dimensionality of excitons
\cite{meier95prl}.
 
\begin{figure*}[t]
\includegraphics[width=2.0\columnwidth]{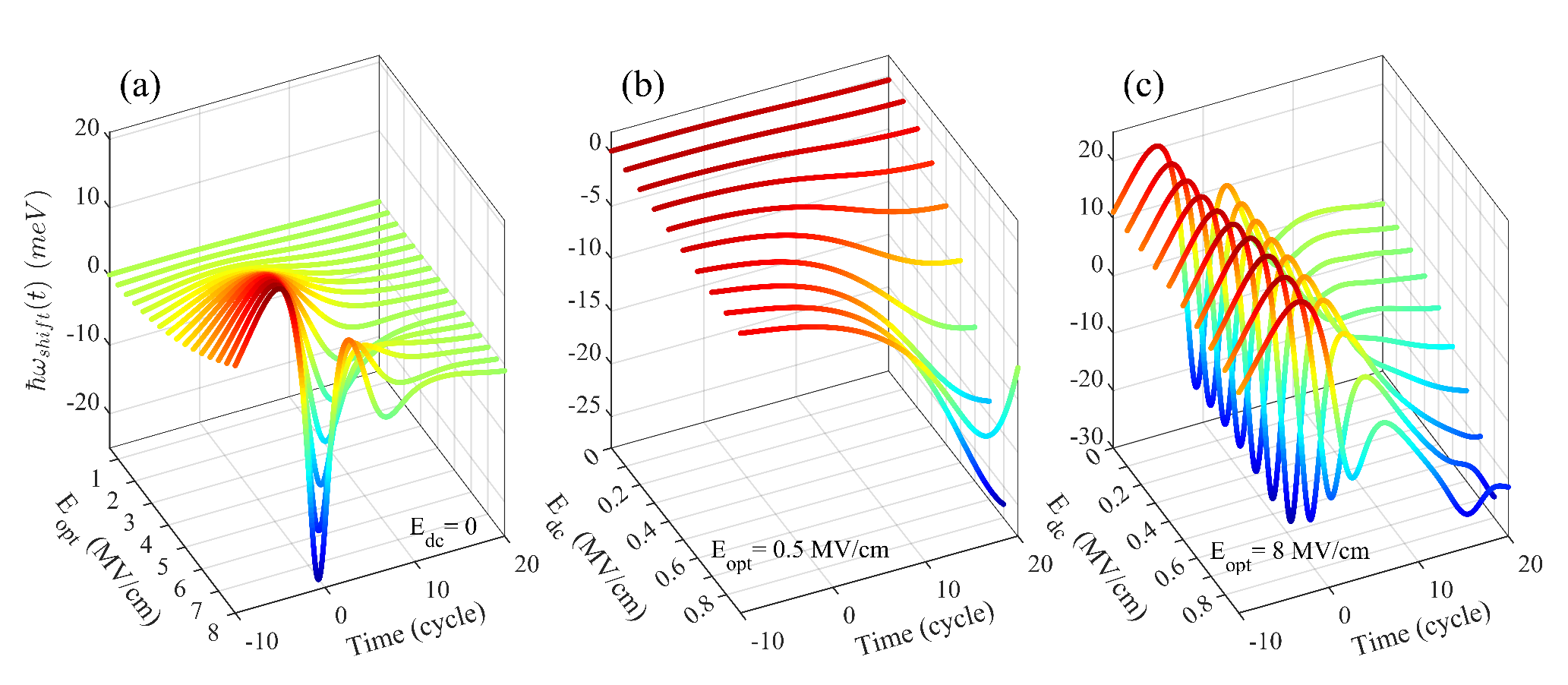}
\caption{$\hbar \omega_{shift}(t)$ of the THG frequency for (a) varying the optical field amplitude with $E_{dc}=0.0~MV/cm$, (b) varying the static field amplitude with $E_{opt}=0.5~MV/cm$, and (c) the same as in (b) but for $E_{opt}=8.0~MV/cm$.
The excitation parameters are the same as those used in Fig.~\ref{fig3}.
The duration of an optical cycle is $6.4~fs$.}
\label{fig6}
\end{figure*}

The energy shifts of the main absorption peaks with respect to the field-free 1s exciton in the presence of both static and cw fields are shown in Fig.~\ref{fig4}(c) and (d). As can be seen in Fig.~\ref{fig4}(c) and (d) for a vanishing dc field, the optical Stark shifts of the 1s exciton can be fitted well by parabolic functions (black dash curves) of the field amplitude with $\alpha=0.6\times10^{-6} e~cm$ for $\hbar\omega_{cw}=1.936~eV/2$ and $\beta=1.35\times10^{-6} e~cm$ for $\hbar\omega_{cw}=1.936~eV/3$.
In the presence of both cw and static fields, we obtain a competition between the dc and the optical Stark shifts which have opposite signs. Therefore the red shift induced by a static field can be compensated or over-compensated by a cw field. For the field amplitudes considered here, the blue shift induced by the cw field dominates for high $E_{cw}$, as exciton ionization limits the achievable dc Stark shifts
and WS localization modifies the nature of the excitons for strong $E_{dc}$.

Next, we analyze the time-dependent frequencies of the coherent SHG and THG polarizations.
For this analysis we filter out frequencies in the range of $1.5\omega_0<\omega<2.5\omega_0$ for SHG and $2.5\omega_0<\omega<3.5\omega_0$ for THG to obtain $\overline{P}(t)$. Then we transform $\overline{P}(t)$ into a rotating frame in order to concentrate on the slowly varying dynamics via $\tilde{P}(t)=\overline{P}(t)e^{-2i\omega_0t}$ for SHG and $\tilde{P}(t)=\overline{P}(t)e^{-3i\omega_0t}$ for THG, respectively. In these frames, subtle changes of the time-dependent phases of the SHG and THG polarizations can be determined via $\phi(t)=Im[log(\tilde{P}(t))]$. If the dynamics is dominated by a single frequency which changes slowly as function of time, the shift of the instantaneous frequency with respect to the field-free 1s exciton resonance is given by $\omega_{shift}(t)=\frac{\partial}{\partial t}\phi(t)$. As shown below, by this procedure we obtain important and detailed information on the static and transient optical Stark shifts of the 1s exciton resonance, i.e., $\delta_{dc}$ and $\delta_{opt}(t)$, and further phenomena.

Figures~\ref{fig5} and ~\ref{fig6} show $\hbar\omega_{shift}(t)$ for the SHG and THG results presented in Figs.~\ref{fig2} and ~\ref{fig3}, respectively.
Fig.~\ref{fig5}(a) demonstrates that for weak optical excitation the SHG starts to oscillate with twice the excitation frequency which is identical to the frequency of the field-free 1s exciton, i.e., $\hbar\omega_{shift}=0$.
With increasing time the instantaneous frequency continuously reduces slightly towards lower frequencies and reaches the position $\delta_{dc}$ of the dc Stark red-shifted exciton after the excitation pulse; this effect is more clearly visible in Fig.~\ref{fig5}(b) for $E_{dc}$ up to about $0.5~MV/cm$.
For slightly larger $E_{opt}$ a temporal increase of $\hbar\omega_{shift}$ becomes visible which follows the envelope of the optical pulse and thus originates from the dynamic optical Stark shift $\delta_{opt}(t)$.
Significant sharp oscillatory changes of $\hbar\omega_{shift}$ appear for stronger optical driving at about $E_{opt}\ge 6~MV/cm$. This dynamics originates from off-resonantly driven Rabi flopping. 
The presence of an associated weak oscillation of the total occupation and sign changes of the polarization (not shown) 
confirms this interpretation.
As can be seen in Fig.~\ref{fig2} for $E_{dc}=0.1~MV/cm$, with increasing excitation intensity the SHG starts to deviate from the perturbative quadratic scaling with $E_{opt}$ which, as shown in Fig.~\ref{fig5}(a), coincides with the start of an incomplete off-resonant Rabi cycle.

Apart from the absence of a dc Stark shift, very similar dynamics is demonstrated in Fig.~\ref{fig6}(a) also for the THG polarization. The deviation of the THG in Fig.~\ref{fig3} from the perturbative scaling with $E_{opt}$ is exactly correlated with the behavior of $\hbar\omega_{shift}(t)$ in Fig.~\ref{fig6}(a) where again off-resonant Rabi oscillations appear for sufficiently strong optical driving. Due to the larger detuning of the optical pulses the number of Rabi cycles is larger for THG than for SHG.

When the optical field is relatively weak, e.g., $E_{opt}=0.5~MV/cm$ shown in Figs.~\ref{fig5}(b) and ~\ref{fig6}(b), the detuning of the exciton with respect to $2 \hbar \omega_0$ or $3 \hbar \omega_0$, respectively, is given by the dc Stark shift $\delta_{dc}$. Therefore, for not too strong dc fields $\hbar\omega_{shift}(t)$ starts at zero, i.e., initially follows the driving, and then continuously shifts red towards $\delta_{dc}$ when the optical pulse is gone. The oscillations that appear in Figs.~\ref{fig5}(b) and ~\ref{fig6}(b) for higher static fields are due to WS localization, i.e., the excitation of more than a single WS exciton. When more than a single transition contributes to the dynamics, the phases of SHG and THG show rapid oscillations due to a coherent quantum-beat like evolution during and also after the excitation. As the WS splitting is proportional to $E_{dc}$ this dynamics becomes more rapid with increasing bias.

For stronger optical fields, e.g., $E_{opt}=8~MV/cm$ shown in Figs.~\ref{fig5}(c) and ~\ref{fig6}(c), the dynamics is dominated by off-resonant Rabi oscillations during the excitation. For not too strong dc fields, $\hbar\omega_{shift}(t)$ ends at $\delta_{dc}$ after the optical pulse is gone. For stronger dc fields, additional oscillatory dynamics appears during and after the excitation which again originates from the excitation of more than a single WS exciton.

As shown above, for the considered excitation conditions several effects contribute to the excitonic SHG and THG. Further insight into the nonlinear dynamics can be obtained by comparing the results obtained by full non-perturbative solutions of the SBE, i.e., Eqs.~(\ref{subeq:7a}) and (\ref{subeq:7b}),  with the perturbative contributions which have been computed using the expansion  with respect to the optical field described in Appendix~\ref{sec:ApdA}. 
Fig.~\ref{fig7}(a) and (b) show the lowest-order perturbative contributions to SHG and THG, i.e., the response up to second- and third-order in the optical field, respectively. The SHG initially rises linearly and then decreases, whereas the THG is initially constant and then decreases as function of $E_{dc}$. For weak optical fields this corresponds to the expected dependencies and the decrease for stronger  $E_{dc}$ originates, as explained above, from the dc Stark shift which detunes the exciton and exciton ionization.
The SHG and THG obtained by solving the full SBE are for small $E_{dc}$ below the lowest-order perturbative results, but for larger $E_{dc}$ above the lowest order perturbative results.
Fig.~\ref{fig7}(a) and (b) thus demonstrates that the quite unexpected results shown in Fig.~\ref{fig2} and Fig.~\ref{fig3}, i.e., the superlinear increase of SHG with $E_{dc}$ and the increase of the THG with $E_{dc}$, clearly originate from higher-order nonlinearities. Thus, whereas for a small bias the lowest and higher-order interfere destructively, for both SHG and THG the interference become constructive for $E_{dc}\ge 0.5MV/cm$. The good agreement between the full results and the perturbative ones which include the four lowest contributing orders demonstrates  furthermore that for the considered peak amplitudes of the optical field $E_{opt}$ we are still in the perturbative limit. This is in fact not the case for more intense optical fields, however, also in this regime the full results remain above the lowest-order perturbative contributions.
It should be noted that in Fig.~\ref{fig7}(a) and (b) the decrease as function of the electrical bias starts at larger $E_{dc}$ for the full result than for the lowest-order perturbative contributions. Besides a transient partial restoration of the resonant excitation of the exciton, when the blue shift $\delta_{opt}(t)$ partly cancels the dc red shift $\delta_{dc}$, also a stabilization of the exciton against ionization in a strong optical field, i.e., a partial dynamic suppression of tunneling \cite{supp_1,supp_2,supp_3}, contributes to this behavior.

\begin{figure}[t]
\includegraphics[width=1.0\columnwidth]{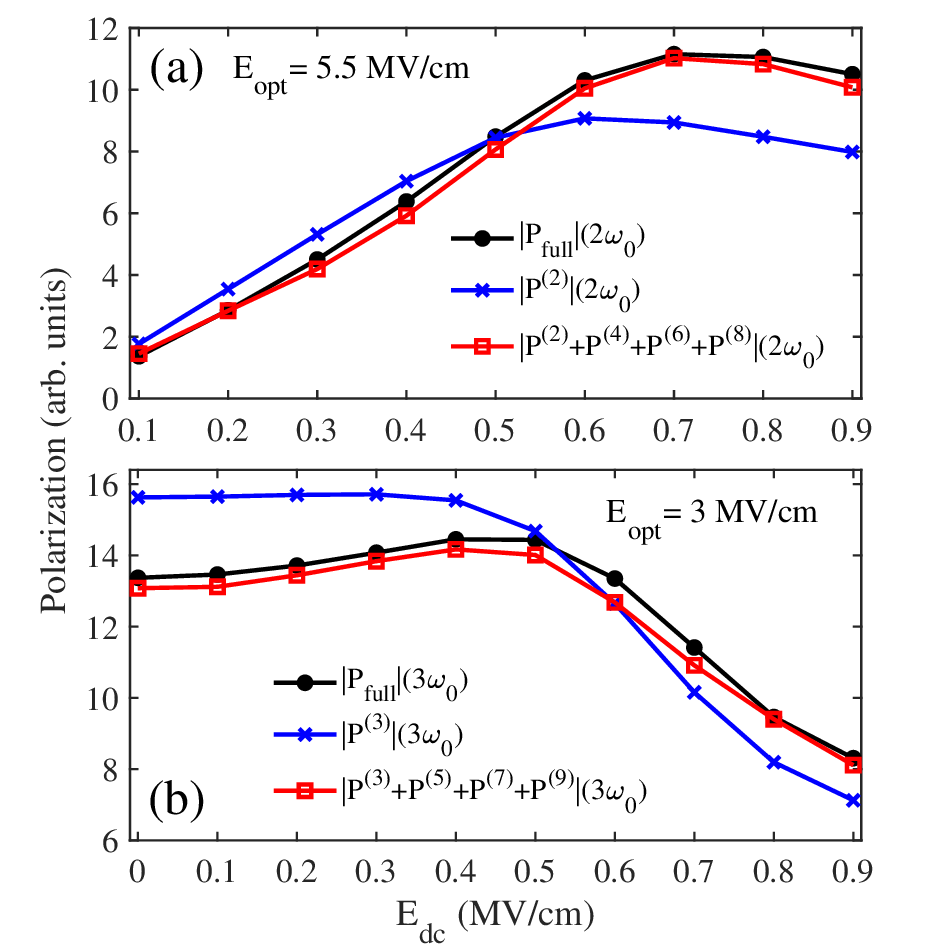}
\caption{
(a) SHG computed as in Fig.~\ref{fig2} by integrating $P(\omega)$ around $2\omega_0$ as function of the amplitude of the static field $E_{dc}$ for $E_{opt}=5.5~MV/cm$.
(b) THG computed as in Fig.~\ref{fig3} by integrating $P(\omega)$ around $3\omega_0$ as function of the amplitude of the static field $E_{dc}$ for $E_{opt}=3~MV/cm$.
For both cases we show the non-perturbative results (black),
the lowest-order perturbative contributions with respect to the optical field, i.e., up to second- and third-order, (blue),
as well as the perturbative expansion  with respect to the optical field including the four lowest contributing orders,  i.e., up to eighth- and ninth-order, (red), respectively.}
\label{fig7}
\end{figure}

\section{\label{sec:conclusions}Conclusions}

In summary, we have numerically solved the SBE including inter- and intraband excitations and the many-body Coulomb interaction to compute resonantly-excited excitonic SHG and THG in the presence of in-plane electric fields. As a model system we investigated the strongly bound intralayer 1s K-exciton of an inversion-symmetric 2H stacked homobilayer of $MoS_2$.
It should be noted, that additional calculations for a generic two-band tight-binding model 
lead to qualitatively similar results which confirms that the obtained dependencies do not depend on details of the model system but are of general nature.

The computed SHG and THG show interesting dependencies on the amplitude of the static field and the peak amplitude of the optical pulse. For weak optical excitation the dependence of the SHG and the THG on the static field shows an expected behavior. The SHG increases  whereas the THG is constant for weak static fields and both decrease for stronger electrical biasing. The decrease is predominantly due to detuning originating from the dc Stark shift of the exciton and exciton ionization.
Very unexpected dependencies on the amplitude of the static field have been demonstrated for stronger optical excitation. We identified a regime in which the THG increases and the SHG increases superlinearly with the amplitude of the static field. Our detailed analysis demonstrates that several effects including static and transient Stark shifts, exciton ionization, Wannier-Stark localization, off-resonant Rabi oscillations, and a modified interference between optical nonlinearities due to the static-field-induced intraband acceleration contribute to the computed signals. 

The understanding gained by our detailed analysis can be used to design new avenues for controlling the strong-field dynamics of systems with strongly bound excitons. By using pulsed Terahertz radiation instead of static fields additional temporal control and higher field strengths might be achievable.

\begin{acknowledgments}
We acknowledge financial support from the Deutsche Forschungsgemeinschaft (DFG) through the Collaborative Research Center TRR 142/3 (Project No. 231447078, Subproject No. A10) and from the National Natural Science Foundation of China (12574310, 12374260). We are grateful for a computation time grant provided by the PC\textsuperscript{2} (Paderborn Center for Parallel Computing). 
\end{acknowledgments}

\appendix

\section{\label{sec:ApdB}Calculation of Coulomb potential}
To describe intralayer excitons in a freestanding bilayer $MoS_2$, we employ the dielectric model for a multilayer structure \cite{Meckbach2018} and compute the quasi-2D intralayer Coulomb potential of a point charge located at the center of the layer. For a geometry of thickness $L=2D$ (D is the layer-to-layer distance in the natural bulk configuration) with dielectric parameters:
\begin{equation}
\epsilon_{\parallel}(z)=\left\{\begin{array}{ll}
     1  & z<-\frac{L}{2}\\
     \epsilon_{\parallel}  & |z|<\frac{L}{2}\\
     1  & z>\frac{L}{2}
\end{array}\right.
,\quad
\epsilon_{\perp}(z)=\left\{\begin{array}{ll}
     1  & z<-\frac{L}{2}\\
     \epsilon_{\perp}  & |z|<\frac{L}{2}\\
     1  & z>\frac{L}{2}
\end{array}\right. .
\end{equation}

The dielectric constant is 1 for the vacuum outside the system and $\epsilon_{\parallel}$ and $\epsilon_{\perp}$ are the background dielectric constants. The bulk dielectric constants $\epsilon_{\perp}$, $\epsilon^B_{\parallel}$ can be obtained from density functional calculations and $\epsilon_{\parallel}$ is determined via $\epsilon_{B}=\epsilon_{\parallel}+\underset{\bm{q}\to0}{\lim} 4\pi e^2\chi_L(\bm{q},\omega)/D$ where $\chi_L(\bm{q},\omega)$ is the longitudinal susceptibility. 

The intralayer Coulomb potential is given by 
\begin{eqnarray}
   &&W_{\bm{q}}(z=\frac{D}{2},z'=\frac{D}{2})=\frac{2\pi}{\kappa q}(1+c_1e^{-\sqrt{\frac{\epsilon_{\parallel}}{\epsilon_{\perp}}}q_{\parallel}D}\nonumber\\
&&+c_1e^{-\sqrt{\frac{\epsilon_{\parallel}}{\epsilon_{\perp}}}q_{\parallel}3D}+2c_2e^{-\sqrt{\frac{\epsilon_{\parallel}}{\epsilon_{\perp}}}q_{\parallel}2L}) , 
\end{eqnarray}
with 
\begin{align}
\kappa&=\sqrt{\epsilon_{\parallel}\epsilon_{\perp}},\nonumber\\
c_1&=\frac{(\kappa+1)(\kappa-1)}{\mathcal{N}},\nonumber\\
c_2&=\frac{(\kappa-1)(\kappa-1)}{\mathcal{N}},\nonumber\\
\mathcal{N}&=(\kappa+1)^2-(\kappa-1)^2e^{-2\sqrt{\frac{\epsilon_{\parallel}}{\epsilon_{\perp}}}q_{\parallel}L} . \nonumber
\end{align}

\section{\label{sec:ApdA}Perturbative expansion of the semiconductor Bloch equations}

For the 2H homobilayer $MoS_2$ considered here the valleys K and K' are equivalent. The interband transition energy is minimal at these valleys and degenerate bands of opposite spin are from upper and lower layers. Here, we consider four bands around the band edge, i.e., the two lowest degenerate conduction and the two highest degenerate valence bands. In this four-band model, the interlayer coupling is rather weak and thus ignored. Throughout this manuscript we consider linearly in-plane polarized optical fields which can be described as a superposition of opposite circularly-polarized fields that induce intralayer transitions of equal strength in both valleys. Using the electron-hole picture, the simplified two-band SBE for the dynamical variables $p_{\bm{k}}=\langle a^{\dagger}_{v\bm{k}}a_{c\bm{k}}\rangle$ and $n_{\bm{k}}=\langle a^{\dagger}_{c\bm{k}}a_{c\bm{k}}\rangle=1-\langle a^{\dagger}_{v\bm{k}}a_{v\bm{k}}\rangle$ read \cite{Haug,Golde2008}
\begin{subequations}
\begin{eqnarray}
\frac{\partial}{\partial t}p_{\bm{k}}=&&-\frac{i}{\hbar}[\varepsilon^c_{\bm{k}}-\varepsilon^v_{\bm{k}}+2\sum_{\bm{k}'}V_{\bm{k}-\bm{k}'}n_{\bm{k}'}-\frac{i\hbar}{T_2})p_{\bm{k}}\nonumber\\
&&-\frac{i}{\hbar}(\bm{d}^{cv}_{\bm{k}}\cdot \bm{E}(t)+\sum_{\bm{k}'}V_{\bm{k}-\bm{k}'}p_{\bm{k}'})(1-2n_{\bm{k}})\nonumber\\
&&+\frac{e}{\hbar}\bm{E}(t)\cdot\nabla_{\bm{k}}p_{\bm{k}},
\label{subeq:1}
\\
\frac{\partial}{\partial t}n_{\bm{k}}=&&\frac{2}{\hbar}Im[ {(\bm{d}^{cv}_{\bm{k}}\cdot \bm{E}(t)+\sum_{\bm{k}'}V_{\bm{k}-\bm{k}'}p_{\bm{k}'})}^{*}{p}_{\bm{k}}]\nonumber\\
&&+\frac{e}{\hbar}\bm{E}(t)\cdot\nabla_{\bm{k}}n_{\bm{k}},
\label{subeq:2}
\end{eqnarray}
\end{subequations}
where $V_{\bm{k}-\bm{k}'}$ describes the intralayer Coulomb interaction. 

The perturbative expansion of the SBE is an effective approach to quantitatively distinguish between inter- and intraband excitation pathways when the excitation intensity is sufficiently weak \cite{Aversa1995,Hannes2019}. 
In the following, we expand the microscopic quantities $X(\bm{k},t)$={$p(\bm{k},t)$, $n_{e/h}(\bm{k},t)$} in powers of the optical field as $X(\bm{k},t)=\sum_{m=0}^{\infty}X^{(m)}(\bm{k},t)$ with $X^{(m)}\propto {E_{opt}}^m$. The initial conditions are $p^{(0)}(\bm{k},t=-\infty)=0$ and ${n}^{(0)}(\bm{k},t=-\infty)=0$. The resulting perturbative set of the two-band SBE reads
\begin{widetext}
\begin{subequations}
\begin{eqnarray}
\frac{\partial}{\partial t}p^{(1)}_{\bm{k}}=&&[-\frac{i}{\hbar}(\varepsilon^c_{\bm{k}}-\varepsilon^v_{\bm{k}})-\frac{1}{T_2}+\frac{e}{\hbar}\bm{E}_{dc}\cdot\nabla_{\bm{k}}]p^{(1)}_{\bm{k}}-2\frac{i}{\hbar}\bm{d}({\bm{k}})\cdot \bm{E}_{dc}n^{(1)}_{\bm{k}}\nonumber\\
&&+ \frac{i}{\hbar}\bm{d}({\bm{k}})\cdot \bm{E}_{opt}(t)+\frac{i}{\hbar}\sum_{\bm{k}'}V_{\bm{k}-\bm{k}'}p^{(1)}_{\bm{k}'},
\label{subeq:1_2} 
\\
\frac{\partial}{\partial t}n^{(1)}_{\bm{k}}=&&\frac{2}{\hbar}Im[\bm{d}^{*}({\bm{k}})\cdot \bm{E}_{dc}{p}^{(1)}_{\bm{k}}]+\frac{e}{\hbar}\bm{E}_{dc}\cdot\nabla_{\bm{k}}n^{(1)}_{\bm{k}},\label{subeq:2_2} 
\\
\frac{\partial}{\partial t}p^{(m)}_{\bm{k}}=&&[-\frac{i}{\hbar}(\varepsilon^c_{\bm{k}}-\varepsilon^v_{\bm{k}})-\frac{1}{T_2}+\frac{e}{\hbar}\bm{E}_{dc}\cdot\nabla_{\bm{k}}]p^{(m)}_{\bm{k}}-2\frac{i}{\hbar}\bm{d}({\bm{k}})\cdot \bm{E}_{dc}n^{(m)}_{\bm{k}}\nonumber\\
&&+\frac{e}{\hbar}\bm{E}_{opt}(t)\cdot\nabla_{\bm{k}}p^{(m-1)}_{\bm{k}}-2\frac{i}{\hbar}\bm{d}({\bm{k}})\cdot \bm{E}_{opt}(t)n^{(m-1)}_{\bm{k}}\nonumber\\
&&+\frac{i}{\hbar}\sum_{\bm{k}'}V_{\bm{k}-\bm{k}'}p^{(m)}_{\bm{k}'}-2\frac{i}{\hbar}\sum_{j\leq m}p^{(j)}_{\bm{k}}\sum_{\bm{k}'}V_{\bm{k}-\bm{k}'}n^{(m-j)}_{\bm{k}'},\label{subeq:3}
\\
\frac{\partial}{\partial t}n^{(m)}_{\bm{k}}=&&\frac{2}{\hbar}Im[\bm{d}^{*}({\bm{k}})\cdot \bm{E}_{dc}{p}^{(m)}_{\bm{k}}]+\frac{e}{\hbar}\bm{E}_{dc}\cdot\nabla_{\bm{k}}n^{(m)}_{\bm{k}}\nonumber\\
&&\frac{2}{\hbar}Im[\bm{d}^{*}({\bm{k}})\cdot \bm{E}_{opt}(t){p}^{(m-1)}_{\bm{k}}]+\frac{e}{\hbar}\bm{E}_{opt}(t)\cdot\nabla_{\bm{k}}n^{(m-1)}_{\bm{k}}\nonumber\\
&&\frac{2}{\hbar}\sum_{j\leq m}Im[{p}^{(j)}_{\bm{k}}\sum_{\bm{k}'}V_{\bm{k}-\bm{k}'}{p^{*}}^{(m-j)}_{\bm{k}'}].\label{subeq:4}
\end{eqnarray}
\end{subequations}
\end{widetext}
The static electric field is considered in the homogeneous part of the respective equations and thus does not couple different orders since we expand with respect to the optical field.
Please note that the static field is switched on smoothly prior to the optical excitation. Furthermore, interband excitations induced by the static field, i.e., the terms $\propto \bm{d}({\bm{k}})\cdot \bm{E}_{dc}$ in Eqs.~(\ref{subeq:1_2})-(\ref{subeq:4})
have a negligible influence on the numerical results that we present.

$E_{dc}$ leads to temporal change of the wave vector $\hbar\dot{\bm{k}}(t)=-e \bm{E}_{dc}$ and couples $p^{(m)}_{\bm{k}}$ and $n^{(m)}_{\bm{k}}$ ($m\geq 1$). $n^{(1)}_{\bm{k}}$ is non-vanishing only in the presence of the static electric field due to interband transitions that could be induced by the static field. Since prior to the excitation the polarization and occupations vanish, in first order the interband excitation induced by the optical field initiates the entire dynamics. Higher orders of $p^{(m)}_{\bm{k}}$ ($m>1$) are generated by intraband excitations via the source term proportional to $p^{(m-1)}_{\bm{k}}$ and interband excitations which are proportional to $n^{(m-1)}_{\bm{k}}$.


\end{document}